\documentclass[12pt]{article}
\usepackage{graphicx}
\usepackage{amsmath,amssymb,amsfonts}
\usepackage{cite}
\usepackage{multirow}
\usepackage{framed}
\usepackage{slashed}
\usepackage{bbm}
\usepackage{array}

\newcount\hour\newcount\minute
        \hour=\time \divide\hour by60 \minute=\time
        {\multiply\hour by60 \global\advance\minute by-\hour}
        \edef\militarytime{\number\hour:\ifnum\minute<10
0\fi\number\minute}

\makeatletter

\@addtoreset{equation}{section}

\def\asymp#1%
{\mathrel{\raisebox{-.4em}{$\widetilde{\scriptstyle #1}$}}}
\makeatother

\newcommand\Ref[1] {Ref.\,\cite{#1}}

\newcommand\eqn[1] {Eq.\,(\ref{#1})}
\newcommand\eqns[2] {Eqs.\,(\ref{#1}) and~(\ref{#2})}

\newcommand\fig[1] {Fig.\,{\ref{#1}}}
\newcommand\sect[1] {Sect.\,{\ref{#1}}}

\newcommand\app[1] {Appendix~\ref{#1}}
\newcommand\tab[1] {Table~\ref{#1}}

\newcommand\subtitle[1] {\noindent{\bf #1}}

\def\beq{\begin{equation}}
\def\eeq{\end{equation}}
\def\bsp#1\esp{\begin{split}#1\end{split}}
\def\bal#1\eal{\begin{align}#1\end{align}}
\def\beeq{\begin{eqnarray}}
\def\eeeq{\end{eqnarray}}

\newcommand\bom[1]     {{\mbox{\boldmath $#1$}}}

\newcommand\bT   {\bom{T}}

\newcommand\rc   {\ensuremath{\mathrm{c}}}
\newcommand\rd   {\ensuremath{\mathrm{d}}}
\newcommand\re   {\ensuremath{\mathrm{e}}}
\newcommand\ri   {\ensuremath{\mathrm{i}}}
\newcommand\rD   {{\mathrm{D}}}
\newcommand\rM   {{\mathrm{M}}}
\newcommand\rL   {{\mathrm{L}}}
\newcommand\rR   {{\mathrm{R}}}
\newcommand\rO   {{\mathrm{O}}}
\newcommand\epeff  {\epsilon_{\mathrm{eff}}}
\newcommand\T[1]{{\mathrm{T}}^{#1}}
\newcommand\Z[1]{{\mathrm{Z}}^{#1}}
\newcommand\rY   {{\mathrm{Y}}}

\newcommand\tS   {\theta_S}
\newcommand\tZ   {\theta_{Z}}
\newcommand\tT   {\theta_{T}}
\newcommand\tW   {\theta_{\mathrm{W}}}
\newcommand\cW   {\cos\tW}
\newcommand\sW   {\sin\tW}
\newcommand\cT   {\cos\tT}
\newcommand\sT   {\sin\tT}
\newcommand\lp   {\ensuremath{\left}}
\newcommand\rp   {\ensuremath{\right}}

\newcommand\cL   {\ensuremath{\mathcal{L}}}
\newcommand\cH   {\ensuremath{\mathcal{H}}}
\newcommand\cS   {\ensuremath{\mathcal{S}}}
\newcommand\cM   {\ensuremath{\mathcal{M}}}
\newcommand\sm   {\ensuremath{\mathrm{SM}}}
\newcommand\lSM  {\ensuremath{\lambda_{\sm}}}

\def\draftdate{\relax}
\def\mda{\relax}
\def\mua{\relax}
\def\mla{\relax}
\def\draft{
\def\thtystars{******************************}
\def\sixtystars{\thtystars\thtystars}
\typeout{}
\typeout{\sixtystars**}
\typeout{* Draft mode!
For final version remove \protect\draft\space in source file *}
\typeout{\sixtystars**}
\typeout{}
\def\draftdate{\today}
\def\mua{\marginpar[\boldmath\hfil$\uparrow$]%
{\boldmath$\uparrow$\hfil}%
\typeout{marginpar: $\uparrow$}\ignorespaces}
\def\mda{\marginpar[\boldmath\hfil$\downarrow$]%
{\boldmath$\downarrow$\hfil}%
\typeout{marginpar: $\downarrow$}\ignorespaces}
\def\mla{\marginpar[\boldmath\hfil$\rightarrow$]%
{\boldmath$\leftarrow $\hfil}%
\typeout{marginpar: $\leftrightarrow$}\ignorespaces}
\overfullrule 5pt
\oddsidemargin -15mm
\marginparwidth 29mm
}

\def\stars{\strut\leaders\hbox{*}\hfill\strut}
\def\starline{\hfil\strut\hfil\hbox to \textwidth {\stars}\hfil}

\begin{document}


\begin{titlepage}
\begin{center}
{\Large \bf
Super-weak force and neutrino masses
}
\end{center}
\par \vspace{2mm}

\begin{center}
{\bf Zolt\'an Tr\'ocs\'anyi},\\[.5em]
{Institute for Theoretical Physics, E\"otv\"os Lor\'and University, \\
P\'azm\'any P\'eter s\'et\'any 1/A, H-1117 Budapest, Hungary and \\
MTA-DE Particle Physics Research Group
}\\
E-mail: {Zoltan.Trocsanyi@cern.ch}
\end{center}

\par \vspace{2mm}
\begin{center}
\today
\end{center}

\par \vspace{2mm}
\begin{center} {\large \bf Abstract} \end{center}
\begin{quote}
\pretolerance 10000
We consider an anomaly free extension of the standard model gauge group
$G_{\rm SM}$ by an abelian group to $G_{\rm SM}\otimes U(1)_Z$. The
condition of anomaly cancellation is known to fix the $Z$-charges of
the particles, but two. We fix one remaining charge by allowing for
all possible Yukawa interactions of the known left handed neutrinos and
new right-handed ones that obtain their masses through interaction
with a new scalar field with spontaneously broken vacuum. We discuss
some of the possible consequences of the model.  Assuming that the new
interaction is responsible for the observed differences between the
standard model prediction for the anomalous magnetic moment of the muon
and its measured value, we constrain the size of the new gauge
coupling, the mass of the new gauge boson and the vacuum expectation
value of the new scalar field.  
\end{quote}

\vspace*{\fill}
\begin{flushleft}
2018
\end{flushleft}
\end{titlepage}
\clearpage

\renewcommand{\thefootnote}{\fnsymbol{footnote}}


\section{Introduction}
\label{sec:intro}

The remarkable experimental success of the standard model of elementary 
particle interaction \cite{Weinberg:1967tq} leaves very little room for the
explanation of the observed deviations from it. This success story has
culminated in the discovery of the Higgs particle
\cite{Aad:2012tfa,Chatrchyan:2012xdj}, which could not have happened
without the immense theoretical input to the design of the accelerator
and the experiments. With this discovery also a new era of particle
physics has arrived as there is no established model that can guide us
to new discoveries.  Therefore, theories that might incorporate the
existing deviations from the standard model are desirable.

The most outstanding experimental observations that cannot be explained
by the standard model are the
(i) abundance of dark matter in the universe;
(ii) non-vanishing neutrino masses;
(iii) leptogenesis;%
\footnote{Baryogenesis can be explained in the standard model provided 
leptogenesis occurs, which is called lepto-baryogenesis.}
(iv) accelerating expansion of the universe, signalling the existence
of dark energy \cite{Tanabashi:2018oca}.%
\footnote{There are numerous other deviations of experimental results
from precision predictions, but to date none has reached the
significance of discovery.} In addition to (i)--(iv), 
(v) inflation in the early universe is also
considered a fairly established fact, although there is no direct proof
for it. All these facts have to be explained by such an extension of the
standard model that respects
(a) the high precision confirmation of the standard model at
collider experiments
(b) and the lack of finding new particles beyond the Higgs boson by
the LHC experiments \cite{ATLAS,CMS}.
There is one more feature of the standard model, the metastability of
vacuum \cite{Bezrukov:2009db,Degrassi:2012ry} that does not necessarily
require new physics, but if new physics exists, it should not worsen
the stability, but possibly push the vacuum to the stability region.

In addition to the experimental success of the standard model, it is 
also highly efficient being based on the concepts of local gauge 
invariance and spontaneous symmetry breaking
\cite{Englert:1964et,Higgs:1964pj}. The only exception of economical
description is the relatively large number of Yukawa couplings of the
fermions needed to explain their masses. The generation of the fermion
masses however, is also highly efficient in the sense that it uses the
same spontaneous symmetry breaking of the scalar field to which all
other particles owe their masses. In this spirit, it is reasonable to
expect that the non-vanishing masses of the neutrinos should be
explained by Yukawa couplings, too.  Also, the choice of the gauge
groups and number of family replications look arbitrary and presently
these are determined by phenomenology only.  

Clearly, the neutrino masses must play a fundamental role in the
possible extensions of the standard model. As the gauge and mass
eigenstates of the neutrinos differ, they must feel a second force to
the gauge interaction. The second force can be a Yukawa coupling
to a scalar. Such explanation of neutrino masses in general requires
the assumption of the existence of right-handed neutrinos and perhaps
a new scalar field.

In the spirit of economy and level of arbitrariness explained above, in
this article we propose an extension of the zoo of particles in the
standard model with three right-handed neutrinos%
and the gauge symmetry of the standard model Lagrangian
$G_\sm = SU(3)_\rc\otimes SU(2)_\rL\otimes U(1)_Y$ to
$G_\sm \otimes U(1)_Z$. Such extensions have already been considered
in the literature extensively (for an incomplete set of popular
examples and their studies see \cite{Schabinger:2005ei,Pospelov:2007mp,
Basso:2008iv}).
In particular, it was shown that the charge assignment of the matter
fields is constrained by the requirement of anomaly cancellations up to
two free charges \cite{Appelquist:2002mw}. To define the model
completely, one has to take a specific choice for these remaining free
charges. In this article we propose that the mechanism for the generation
of neutrino masses fixes the values of the $U(1)_Z$ charges up to
an overall scale that can be embedded in the $U(1)_Z$ coupling.

The difference between our proposal and existing studies is two-fold. 
The model proposed here introduces a new force along the same
principles as the known forces are included in the standard model: all
renormalizable terms that are allowed by the underlying gauge symmetry
are present, but no other symmetry than the extra $U(1)_Z$ is assumed.
Our primary goal is not the prediction of new observable phenomena at
collider experiments, but first focus only on the unexplained phenomena
(i--iv), with respecting the observations (a) and (b). As the
deviations from the standard model are related to the intensity and
cosmic frontiers of particle physics, we assume that the new $U(1)_Z$
interaction is secluded from the standard model by a small coupling. 
Thus we propose the model in a region of the parameter space that has
received little attention before. 


\section{Definition of the model}
\label{sec:model}


\subsection{Fermion sector}

We consider the usual three fermion families of the standard model
extended with one right-handed Dirac neutrino in each family.%
\footnote{We find natural to assume one extra neutrino in each family
although known observations do not exclude other possibilities.}
We introduce the notation
\beq
\bsp
\psi_{q,1}^f &=
\lp(\!\!\begin{array}{c}
 U^f \\
 D^f
\end{array}\!\!\rp)_\rL
\qquad\psi_{q,2}^f = U_{\rR}^f
\,,\qquad
\psi_{q,3}^f = D_{\rR}^f
\\[2mm]
\psi_{l,1}^f &=
\lp(\!\!\begin{array}{c}
 \nu^f \\
 \ell^f
\end{array}\!\!\rp)_\rL
\qquad\psi_{l,2}^f = \nu_{\rR}^f
\,,\qquad
\psi_{l,3}^f = \ell_{\rR}^f
\label{eq:psi_j}
\esp
\eeq
for the chiral quark fields $\psi_q$ and chiral lepton fields $\psi_l$.
In \eqn{eq:psi_j} L and R denote the left and right-handed projections
of the same field,
\footnote{The Weyl spinors of $\nu_\rL$ and $\nu_\rR$ can be embedded
into {\em different} Dirac spinors, leading to Majorana neutrinos,
without essential changes in the model.  However, the negative results
of the experiments searching for neutrinoless double $\beta$-decay make
the Majorana nature of neutrinos increasingly unlikely.}
\beq
\psi_{\rL/\rR}\equiv \psi_\mp =
\frac{1}{2}\lp(1\mp\gamma_{5}\rp)\psi \equiv P_{\rL/\rR} \psi
\,.
\label{eq:LRprojections}
\eeq
Then the field content in family $f$ ($f=1$, 2 or 3) consists of two
quarks, $U_f$, $D_f$, a neutrino $\nu_f$ and a charged lepton $\ell_f$
where $U_f$ is the generic notation for the u-type quarks $U_1 = $ u,
$U_2 = $ c, $U_3 = $ t, while $D_f$ is that for d-type quarks, $D_1=$ d,
$D_2 = $ s and $D_3 = $ b.  The charged leptons $\ell_f$ can be $\ell_1
= e$, $\ell_2 = \mu$ or $\ell_3 = \tau$ and $\nu_f$ are the
corresponding neutrinos, $\nu_1=\nu_e$, $\nu_2=\nu_\mu$, $\nu_3=\nu_\tau$.

For a matrix $U\in G_{\sm}\otimes U(1)_Z$ the three generic fields in
\eqn{eq:psi_j} transform as 
\beq
\bsp
U\psi_{1}\lp(x\rp) &=
\re^{\ri\bT\cdot\bom{\alpha}\lp(x\rp)}
\,\re^{\ri\,y_{1}\beta\lp(x\rp)}
\,\re^{\ri\,z_{1}\zeta\lp(x\rp)}
\,\psi_{1}(x)
\quad \textrm{where}\quad\bT=\frac12 (\tau_1,\tau_2,\tau_3)
\\
U\psi_{j}\lp(x\rp) &=
\qquad\qquad
\re^{\ri\,y_{j}\beta\lp(x\rp)}
\re^{\ri\,z_{j}\zeta\lp(x\rp)}\psi_{j}(x)
\quad \textrm{where} \quad\,j=2,3
\esp
\eeq
and $\bom{\alpha} = (\alpha_1,\alpha_2,\alpha_3)$, with $\alpha_i$, $\beta$, 
$\zeta\in\mathbb{R}$. The matrices $\tau_{i}$ are the Pauli matrices,
$y_j$ is the hypercharge, while $z_j$ denotes the $Z$-charge of the
field $\psi_j$. There is a lot of freedom how to choose the $Z$-charges.
In this article we make two assumptions that fix these completely.
The first is that the charges do not depend on the families, which
is also the case in the standard model.%
\footnote{Several recent observations hint at violation of lepton flavour
universality, which may be taken into account in our model by choosing
family dependent $Z$-charges. However, those results are controversial
at present, so we neglect them.}
With this assumption the assignment for the $Z$-charges of the fermions
can be expressed using two free numbers $Z_1$ and $Z_2$ of the $U$
quark fields if we want a model free of gauge and gravity anomalies.
The rest of the charges must take values as given in \tab{tab:charges}
\cite{Appelquist:2002mw}.
\begin{table}
\newcommand\zj[1]{$\frac{#1}6$}
\begin{center}
\caption{Assignments for the representations (for $SU(N)$) and charges
(for $U(1)$) of fermion and scalar fields of the complete model. The
charges $y_j$ denote the eigenvalue of $Y/2$, with $Y$ being the
hypercharge operator and $z_j$ denote the supercharges of the fields
$\psi_j$ of \eqn{eq:psi_j} ($j=1$, 2, 3).
The right-handed Dirac neutrinos $\nu_{\rR}$ are sterile under the
$G_{\rm SM}$ group. The sixth column gives a particular realization of
the $U(1)_Z$ charges, motivated below, and the last one is added for
later convenience.} 
\label{tab:charges}.
\begin{tabular}{lccrr|rc}
\hline
\hline
field & $SU(3)_\rc$ & $SU(2)_\rL$ & $y_j$  & $z_j$&$z_j$ &$r_j=z_j/z_\phi-y_j$\\
\hline
$U_\rL$, $D_\rL$     & 3 & 2 & $\frac16$ &    $Z_1$    & $\frac16$ &~~0  \\[2mm]
$U_\rR$              & 3 & 1 & $\frac23$ &    $Z_2$    & $\frac76$ &~~$\frac12$  \\[2mm]
$D_\rR$              & 3 & 1 &$-\frac13$ & $2 Z_1-Z_2$ &$-\frac56$ &$-\frac12$ \\[2mm]
$\nu_\rL$, $\ell_\rL$& 1 & 2 &$-\frac12$ & $-3 Z_1$    &$-\frac12$ &~~0  \\[2mm]
$\nu_\rR$            & 1 & 1 & 0         & $Z_2-4 Z_1$ & $\frac12$ &~~$\frac12$  \\[2mm]
$\ell_\rR$           & 1 & 1 & $-1$      & $-2Z_1-Z_2$ &$-\frac32$ &$-\frac12$ \\[2mm]
$\phi$               & 1 & 2 & $\frac12$ & $z_\phi$    &  1        &~~$\frac12$  \\[2mm]
$\chi$               & 1 & 1 & 0         & $z_\chi$    &$-1$       &$-1$ \\
\hline
\hline
\end{tabular}
\end{center}
\end{table}

The Dirac Lagrangian summed over the family replications,
\beq
\bsp
\cL_\rD &= \ri\,\sum_{f=1}^{3}\sum_{j=1}^{3}\Big(
  \overline{\psi}^f_{q,j}(x)\slashed{D}_{j}\psi_{q,j}^f(x)
+ \overline{\psi}^f_{l,j}(x)\slashed{D}_{j}\psi_{l,j}^f(x)
\Big)
\,,
\\[2mm]
D^{\mu}_{j} &= \partial^{\mu}
+ \ri g_\rL\,\bT\cdot\bom{W}^{\mu}
+ \ri g_Y\,y_{j}B^{\mu}
+ \ri g_Z\,z_{j}Z^{\mu}
\label{eq:LD}
\esp
\eeq
is invariant under local $G=G_{\rm SM}\otimes U(1)_Z$ gauge
transformations, provided the five gauge fields introduced in the
covariant derivative transform as
\beq
\bsp
\bT\cdot\bom{W}^{\mu}(x)&\stackrel{G}{\longrightarrow}
\bT\cdot\bom{W}^{\prime\mu}(x)=
U(x)\,\bT\cdot\bom{W}^{\mu}(x)\,U^{\dag}(x)
+\frac{\ri}{g_\rL}\lp[\partial^{\mu}\,U(x)\rp]U^{\dag}(x)
\\
B^{\mu}&\stackrel{G}{\longrightarrow}
B^{\prime\mu}(x)=B^{\mu}(x)-\frac{1}{g_Y}\,\partial^{\mu}\beta(x)
\\
Z^{\mu}&\stackrel{G}{\longrightarrow}
Z^{\prime\mu}(x)=Z^{\mu}(x)-\frac{1}{g_Z}\,\partial^{\mu}\zeta(x)
\label{eq:WBXtransformations}
\esp
\eeq
where
$U(x)=\exp\lp[\ri\bT\cdot\bom{\alpha}\lp(x\rp)\rp]$.
The gauge invariant kinetic term for these vector fields is
\begin{equation}
\cL_{B,Z,W}=
-\frac{1}{4}B_{\mu\nu}B^{\mu\nu}
-\frac{1}{4}Z_{\mu\nu}Z^{\mu\nu}
-\frac{1}{4}\bom{W}_{\mu\nu}\cdot\bom{W}^{\mu\nu},
\label{eq:LBZW}
\end{equation}
with
$B_{\mu\nu} = \partial_{\mu}B_{\nu}-\partial_{\nu}B_{\mu}
\equiv\partial_{[\mu}B_{\nu]}$, $Z_{\mu\nu} = \partial_{[\mu}Z_{\nu]}$
and
$\bom{W}_{\mu\nu} = \partial_{[\mu}\bom{W}_{\nu]}
-g\,\bom{W}_{\mu}\times\bom{W}_{\nu}$. The field strength
$\bT\cdot\bom{W}_{\mu\nu}$ transforms covariantly under $G$
transformations, $\bT\cdot\bom{W}_{\mu\nu}\stackrel{G}{\longrightarrow}
U(x)\,\bT\cdot\bom{W}_{\mu\nu}\,U^{\dag}(x)$, but
$B_{\mu\nu}$ and $Z_{\mu\nu}$ are invariant, hence a kinetic mixing
term of the $U(1)$ fields is also allowed by gauge invariance:
\beq
- \frac{\epsilon}{2} B_{\mu \nu} Z^{\mu \nu}\,.
\label{eq:kinetic-mixing}
\eeq
We can get rid of this mixing term by redefining the $U(1)$ fields using
the transformation
\beq
\lp(\!\!\begin{array}{c}
 B_\mu' \\
 Z_\mu'
\end{array}\!\!\rp) =
\lp(\!\!\begin{array}{cc}
 1 & \sin\tZ \\
 0 & \cos\tZ
\end{array}\!\!\rp)
\lp(\!\!\begin{array}{c}
 B_\mu \\
 Z_\mu
\end{array}\!\!\rp)
\,,\qquad\sin\tZ = \epsilon
\,.
\eeq
In terms of the redefined fields, the covariant derivative becomes
\beq
D^{\mu}_j= \partial^{\mu}
+\ri g_\rL\,\bT\cdot\bom{W}^{\mu}+\ri g_Y\,y_{j}B^{\prime\mu}
+\ri (g_Z'\,z_{j} - g_Y'\,y_{j})Z^{\prime\mu}
\label{eq:cov-dev}
\eeq
where $g_Y' = g_Y \tan\tZ = \epsilon g_Y + \rO(\epsilon^3)$
and $g_Z' = g_Z/\cos\tZ = g_Z  + \rO(\epsilon^2)$. Thus the effect
of the kinetic mixing is to change the couplings of the matter fields
to the vector field $Z^\mu$. Note that we cannot immediately combine the
coupling factor $(g_Z'\,z_{j} - g_Y'\,y_{j})$ into a single product of
a coupling and a charge. We shall discuss this issue further below. 

Gauge symmetry forbids mass terms for gauge bosons. Fermion masses must
also be absent because
\[m\,\bar{\psi}\psi =
m\,\bar{\psi}_\rL\psi_\rR +m\,\bar{\psi}_\rR \psi_\rL,\]
but the $\psi_\rL$, $\psi_\rR $ fields transform differently under $G$.
Thus the $G$-invariant Lagrangian describes massless fields in
contradiction to observation.


\subsection{Scalar sector}

To solve the puzzle of missing masses we proceed similarly as in the
standard model, but in addition to the usual Brout-Englert-Higgs (BEH)
field $\phi$ that is an $SU(2)_\rL$-doublet
\begin{equation}
\phi=\lp(\!\!\begin{array}{c}
               \phi^{+} \\
               \phi^{0}
             \end{array}\!\!\rp) = 
\frac{1}{\sqrt{2}}
\lp(\!\!\begin{array}{c}
          \phi_{1}+\ri\phi_{2} \\
          \phi_{3}+\ri\phi_{4}
        \end{array}\!\!
\rp)
\,,
\end{equation}
we also introduce another complex scalar $\chi$ that transforms as a
singlet under $G_{\rm SM}$ transformations.  The gauge invariant
Lagrangian of the scalar fields is
\beq
\cL_{\phi,\chi} =
  [D_{\phi\,\mu} \phi]^* D_{\phi}^{\mu} \phi
+ [D_{\chi\,\mu} \chi]^* D_{\chi}^{\mu} \chi
- V(\phi,\chi)
\label{eq:Lphichi}
\eeq
where the covariant derivative for the scalar $s$ ($s=\phi$, $\chi$) is
\beq
D_{s}^{\mu} = \partial^{\mu}
+\ri g_\rL\,\bT\cdot\bom{W}^{\mu}+\ri g_Y\,y_{s}B^{\prime\mu}
+\ri (g_Z'\,z_{s} - g_Y'\,y_{s})Z^{\prime\mu}
\label{eq:cov-dev-scalar}
\eeq
and the potential energy
\beq
V(\phi,\chi) = V_0 - \mu_\phi^2 |\phi|^2 - \mu_\chi^2 |\chi|^2
+ \lp(|\phi|^2, |\chi|^2\rp)
\lp(\!\!\begin{array}{cc}
 \lambda_\phi & \frac{\lambda}2 \\ \frac{\lambda}2 & \lambda_\chi 
\end{array}\!\!\rp) 
\lp(\!\!\begin{array}{c}
|\phi|^2 \\ |\chi|^2
\end{array}\!\!\rp)\,,
\label{eq:V}
\eeq
in addition to the usual quartic terms, introduces a coupling term
$-\lambda |\phi|^2 |\chi|^2$ of the scalar fields in the Lagrangian.
For the doublet $|\phi|$ denotes the length $\sqrt{|\phi^+|^2 + |\phi^0|^2}$.
The value of the additive constant $V_0$ is irrelevant for particle
dynamics, but may be relevant for inflationary scenarios, hence we
allow for its non-vanishing value. In order that this potential energy
be bounded from below, we have to require the positivity of the
self-couplings, $\lambda_\phi$, $\lambda_\chi>0$. The eigenvalues
of the coupling matrix are
\beq
\lambda_\pm = \frac12 \left(\lambda_\phi+\lambda_\chi
\pm \sqrt{(\lambda_\phi-\lambda_\chi)^2 + \lambda^2}\right)
\,,
\eeq
while the corresponding un-normalized eigenvectors are
\beq
u^{(+)} = \lp(\!\!\begin{array}{c}
\frac{2}{\lambda} (\lambda_+ - \lambda_\chi) \\ 1
\end{array}\!\!\rp)
\quad\textrm{and}\quad
u^{(-)} = \lp(\!\!\begin{array}{c}
\frac{2}{\lambda} (\lambda_- - \lambda_\chi) \\ 1
\end{array}\!\!\rp)
\,.
\eeq
As $\lambda_+>0$ and $\lambda_-<\lambda_+$, in the physical region the
potential can be unbounded from below only if $\lambda_-<0$ and
$u^{(-)}$ points into the first quadrant, which may occur only when
$\lambda<0$. In this case, to ensure that the potential is bounded from
below, one also has to require that the coupling matrix be positive
definite, which translates into the condition
\beq
4 \lambda_\phi \lambda_\chi - \lambda^2 > 0
\,.
\label{eq:positivity}
\eeq
With these conditions satisfied, we can find the minimum of the
potential energy at field values $\phi= v/\sqrt{2}$ and
$\chi = w/\sqrt{2}$ where the vacuum expectation values (VEVs) are
\beq
v = \sqrt{2} \sqrt{\frac{2\lambda_\chi \mu_\phi^2 - \lambda \mu_\chi^2}
{4 \lambda_\phi \lambda_\chi - \lambda^2}}
\,,\qquad
w = \sqrt{2} \sqrt{\frac{2\lambda_\phi \mu_\chi^2 - \lambda \mu_\phi^2}
{4 \lambda_\phi \lambda_\chi - \lambda^2}}
\,.
\label{eq:VEVs}
\eeq
Using the VEVs, we can express the quadratic couplings as
\beq
\mu_\phi^2 = \lambda_\phi v^2 + \frac{\lambda}{2} w^2
\,,\qquad
\mu_\chi^2 = \lambda_\chi w^2 + \frac{\lambda}{2} v^2
\,,
\label{eq:scalarmasses}
\eeq
so those are both positive if $\lambda > 0$. If $\lambda < 0$, the
constraint (\ref{eq:positivity}) ensures that the denominators of the
VEVs in \eqn{eq:VEVs} are positive, so the VEVs have non-vanishing real
values only if
\beq
2\lambda_\chi \mu_\phi^2 - \lambda \mu_\chi^2 > 0
\quad\text{and}\quad
2\lambda_\phi \mu_\chi^2 - \lambda \mu_\phi^2 > 0
\label{eq:muXconditions}
\eeq
simultaneously, which can be satisfied if at most one of the quadratic
couplings is smaller than zero. We summarize the possible cases for the
signs of the couplings in \tab{tab:scalarcouplings}.
\begin{table}
\newcommand\zj[1]{$\frac{#1}6$}
\begin{center}
\caption{Possible signs of the couplings in the scalar potential 
$V(\phi,\chi)$ in order to have two non-vanishing real VEVs.
$\Theta$ is the step function, $\Theta(x)=1$ if $x>0$ and 0 if $x<0$.}
\label{tab:scalarcouplings}.
\begin{tabular}{c|ccc|cc}
\hline
\hline
$\Theta(\lambda)$ &
$\Theta(\lambda_\phi)\!$ & $\!\Theta(\lambda_\chi)$ &
$\Theta(4 \lambda_\phi \lambda_\chi - \lambda^2)$ &
$\Theta(\mu_\phi^2)\,\Theta(\mu_\chi^2)$ &
$\Theta(2\lambda_\chi \mu_\phi^2 - \lambda \mu_\chi^2)
 \Theta(2\lambda_\phi \mu_\chi^2 - \lambda \mu_\phi^2)$\\
\hline
1 & 1 & 1 & unconstrained & 1 &  unconstrained \\[2mm]
0 & 1 & 1 & 1             & 1 &  unconstrained \\
0 & 1 & 1 & 1             & 0 & 1  \\
\hline
\hline
\end{tabular}
\end{center}
\end{table}

After spontaneous symmetry breaking of $G \to SU(3)_\rc\otimes U(1)_Q$,
we use the following convenient parametrization for the
scalar fields: 
\begin{equation}
\phi=\frac{1}{\sqrt{2}}\,\re^{\ri\bT\cdot\bom{\xi}(x)/v}
\lp(\!\!\begin{array}{c} 0 \\ v+h'(x) \end{array}\!\!\rp)
\quad\mbox{and}\quad
\chi(x) =
\frac{1}{\sqrt{2}}\,\re^{\ri\eta(x)/w}\big(w + s'(x)\big)
\,.
\label{eq:BEHparametrization}
\end{equation}
We can use the gauge invariance of the model to choose the unitary
gauge when
\beq
\phi'(x)=
\frac{1}{\sqrt{2}} \lp(\!\!\begin{array}{c} 0 \\ v+h'(x) \end{array}\!\!\rp)
\quad\mbox{and}\quad
\chi'(x) = \frac{1}{\sqrt{2}}\big(w + s'(x)\big)
\eeq
and the vector fields are transformed according to 
\eqn{eq:WBXtransformations}.
With this gauge choice, the scalar kinetic term contains quadratic
terms of the gauge fields from which one can identify mass parameters
of the massive standard model gauge bosons proportional to the
vacuum expectation value $v$ of the BEH field and also that of a
massive vector boson $Z^{\prime\mu}$ proportional to $w$.

We can diagonalize the mass matrix (quadratic terms) of the two real
scalars ($h'$ and $s'$) by the rotation
\beq
\lp(\!\!\begin{array}{c}
 h \\
 s
\end{array}\!\!\rp) =
\lp(\!\!\begin{array}{cr}
 \cos\tS & -\sin\tS \\
 \sin\tS &  \cos\tS
\end{array}\!\!\rp)
\lp(\!\!\begin{array}{c}
 h' \\
 s'
\end{array}\!\!\rp)
\eeq
where for the scalar mixing angle $\tS \in (-\frac\pi4,\frac\pi4)$ we find
\beq
\sin(2\tS) = - \frac{\lambda v w}
{\sqrt{(\lambda_\phi v^2 - \lambda_\chi w^2)^2 + (\lambda v w)^2}}
\,.
\eeq
The masses of the mass eigenstates $h$ and $s$ are
\beq
M_{h/H} = \lp(\lambda_\phi v^2 + \lambda_\chi w^2
\mp \sqrt{(\lambda_\phi v^2 - \lambda_\chi w^2)^2 + 
(\lambda v w)^2}\rp)^{1/2}
\label{eq:Mhs}
\eeq
where $M_h \leq M_H$ by convention. At this point either $h$ or $H$ can
be the standard model Higgs boson. A more detailed analysis of this
scalar sector but within a different $U(1)_Z$ model can be found in
\Ref{Duch:2015jta} and for the present model in \Ref{Peli:2019xwv}.

\subsection{Fermion masses}
\label{ssec:Yukawas}

We already discussed that explicit mass terms of fermions would break
$SU(2)_\rL\otimes U(1)_Y$ invariance. However, we can introduce 
gauge-invariant fermion-scalar Yukawa interactions%
\footnote{We distinguish the hypercharge $Y$ from the index referring
to Yukawa terms using different type of letters.}
\beq
\cL_{\rY} =
-\lp[c_{D}
\bar{Q}_\rL\cdot\phi\:D_\rR 
+c_{U}
\bar{Q}_\rL\cdot\tilde{\phi}\:U_\rR 
+c_{\ell}
\bar{L}_\rL\cdot\phi\:\ell_\rR\rp]
+ \mathrm{h.c.}
\label{eq:LY}
\eeq
where h.c.\ means hermitian conjugate terms
and the parameters $c_{D},\,c_{U},\,c_{\ell}$ are called Yukawa
couplings that are matrices in family indices and summation over the
families is understood implicitly. The dot product abbreviates scalar 
products of $SU(2)$ doublets:
\beq
\bar{Q}_\rL\cdot\phi \equiv
\lp(\bar{U},\,\bar{D}\rp)_\rL
\lp(\!\!\begin{array}{c}
  \phi^{\lp(+\rp)} \\
  \phi^{\lp(0\rp)}
\end{array}\!\!\rp)
\,,\quad
\bar{Q}_\rL\cdot\tilde{\phi} \equiv
\lp(\bar{U},\,\bar{D}\rp)_\rL
\lp(\!\!\begin{array}{r}
  \phi^{\lp(0\rp)\,*} \\
  -\phi^{\lp(+\rp)\,*}
\end{array}\!\!\rp)
\eeq
and $\bar{L} \equiv \lp(\bar{\nu}_{\ell},\,\bar{\ell}\rp)$.
The $Z$-charge of the BEH field is
constrained by $U(1)_Z$ invariance of the Yukawa terms to
$z_\phi = Z_2 - Z_1$, which works simultaneously for all three terms.

After spontaneous symmetry breaking and fixing the unitary gauge, this
Yukawa Lagrangian becomes
\begin{equation}
\cL_{\rY} =
-\frac{1}{\sqrt{2}}\lp(v+h(x)\rp)
\lp[c_{D}\,\bar{D}_\rL D_\rR
  + c_{U}\,\bar{U}_\rL U_\rR
  + c_{\ell}\,\bar{\ell}_\rL \ell_\rR \rp]
+{\rm h.c.}
\label{eq:LYcharged}
\end{equation}
We see that there are mass terms with mass matrices
$M_{i}=\frac{c_{i}v}{\sqrt{2}}$ where $i=D$, $U$, $\ell$:
\begin{equation}
\cL_{\rY} =
- \lp(1+\frac{h(x)}{v}\rp)
\lp[\bar{D}_\rL\,M_{D}\,D_\rR
  + \bar{U}_\rL\,M_{U}\,U_\rR
  + \bar{\ell}_\rL\,M_{\ell}\,\ell_\rR\rp]
+{\rm h.c.}
\label{eq:chargedmasses}
\end{equation}
The general complex matrices $M_{i}$ can be diagonalized employing
bi-unitary transformations. The diagonal elements on the basis of mass
eigenstates provide the mass parameters of the fermions. Due to the
bi-unitary transformation the left and right-handed components of the
fermion field are different linear combinations of the mass eigenstates.

The neutrino oscillation experiments suggest non-vanishing neutrino
masses and the weak and mass eigenstates of the left-handed neutrinos
do not coincide. In principle, the charge assignment of our model
allows for the following gauge invariant Yukawa terms of dimension four
operators for the neutrinos
\beq
\cL^\nu_{\rY} =
-\sum_{i,j}\left(
(c_{\nu})_{ij}
\bar{L}_{i,\rL}\cdot\tilde{\phi}\:\nu_{j,\rR} 
+ \frac12 (c_\rR)_{ij}\,\overline{\nu_{i,\rR}^c} \nu_{j,\rR}\,\chi
\right)
+ {\rm h.c.}
\label{eq:nuYukawa}
\eeq
for arbitrary values of $Z_1$ and $Z_2$ if the superscript $c$ denotes
the charge conjugate of the field, $\nu^c = -\ri \gamma_2 \nu^*$ and
the $Z$-charge of the right-handed neutrinos and the new scalar satisfy
the relation $z_\chi = -2 z_{\nu_\rR}$. There are two natural choices
to fix the $Z$-charges: (i) the left- and right-handed neutrinos have
the same charge, or (ii) those have opposite charges. In the first case
we have
\beq
Z_2-4 Z_1 = -3 Z_1
\,,
\eeq
which is solved by $Z_1 = Z_2$ and it leads to the charge assignment of
the $U(1)_{B-L}$ extension of the standard model, studied in detail
(see for instance, \cite{Basso:2011hn} and references therein).
In the second case
\beq
Z_2-4 Z_1 = 3 Z_1
\,,
\label{eq:Z2-Z1}
\eeq
which is solved by $Z_1 = Z_2/7$. As the overall scale of the
$Z$-charges depends only on the value of the gauge coupling $g_Z'$, we
set $Z_2$ freely. For instance, choosing $Z_2=7/6$ implies $Z_1 = 1/6$
and the $Z$-charge of the BEH scalar is 
\beq
z_\phi = 1
\,,
\label{eq:zphi}
\eeq
while that of the new scalar is
\beq
z_\chi = -1 = -z_\phi
\,.
\eeq

While we cannot exclude the infinitely many cases when the magnitudes
of $Z$-charges of the left- and right-handed neutrinos differ, we find
natural to assume that \eqn{eq:Z2-Z1} is valid. The corresponding
$Z$-charges are given explicitly in the sixth column of
\tab{tab:charges}.  

After the spontaneous symmetry breaking of the vacuum of the scalar
fields \eqn{eq:nuYukawa} leads to the following mass terms for the
neutrinos:
\beq
\cL^\nu_{\rY} =
-\frac12 \sum_{i,j}\Bigg[
\lp(\overline{\nu_{\rL}},\,\overline{\nu^c_{\rR}}\rp)_i
M(h,s)_{ij}
\lp(\!\!\begin{array}{r}
  \nu^c_{\rL} \\
  \nu_\rR
\end{array}\!\!\rp)_j
+ {\rm h.c.} \Bigg]
\label{eq:numasses1}
\eeq
where
\beq
M(h,s)_{ij} = 
\lp(\!\!\begin{array}{cc}
  0 & m_\rD \left(1+\frac{h}{v}\right)\\[2mm]
  m_\rD \left(1+\frac{h}{v}\right) & M_\rM \left(1+\frac{s}{w}\right)
\end{array}\!\!\rp)_{ij}
\,,
\label{eq:massmatrix1}
\eeq
with complex $m_\rD$ and real $M_\rM$ being symmetric $3\times 3$ matrices,
so $M(0,0)$ is a complex symmetric $6\times 6$ matrix.
The diagonal elements of the mass matrix $M(0,0)$ provide Majorana mass terms
for the left-handed and right-handed neutrinos.  Thus we conclude that
the model predicts {\em vanishing masses of the left-handed neutrinos}
at the fundamental level.

The off-diagonal elements represent interaction terms that look
formally like Dirac mass terms,
$- \sum_{i,j} \overline{\nu_{i,\rL}} (m_\rD)_{ij} \nu_{j,\rR} + $ h.c.
After spontaneous symmetry breaking the quantum numbers of the
particles $\nu^c_{i,\rL}$ and $\nu_{i,\rR}$ are identical, hence they
can mix. Thus the propagating states will be a mixture of the left- and
right-handed neutrinos. Those states can be obtained by the diagonalization
of the full matrix $M(0,0)$, for which a possible parametrization is
given for instance in \Ref{Blennow:2011vn}.  

In order to understand the structure of the matrix $M(0,0)$ better, we
first diagonalize the matrices $m_\rD$ and $M_\rM$ separately by a 
unitary transformation and an orthogonal one. Defining
\beq
\nu'_{\rL,i} = \sum_j (U_\rL)_{ij} \nu_{\rL,j}
\quad\textrm{and}\quad
\nu'_{\rR,i} = \sum_j (O_\rR)_{ij} \nu_{\rR,j}
\,,
\eeq
we can rewrite the neutrino Yukawa Lagrangian as
\beq
\cL^\nu_{\rY} =
-\frac12 \sum_{i,j}\Bigg[
\lp(\overline{\nu'_{\rL}},\,\overline{\nu^{'c}_{\rR}}\rp)_i
M'(h,s)_{ij}
\lp(\!\!\begin{array}{r}
  \nu^{'c}_{\rL} \\
  \nu'_\rR
\end{array}\!\!\rp)_j
+ {\rm h.c.} \Bigg]
\label{eq:numasses2}
\eeq
where
\beq
M'(h,s) = 
\lp(\!\!\begin{array}{cc}
  0 & m V \left(1+\frac{h}{v}\right)\\[2mm]
  V^\dag m \left(1+\frac{h}{v}\right) & M \left(1+\frac{s}{w}\right)
\end{array}\!\!\rp)
\,.
\label{eq:massmatrix2}
\eeq
In \eqn{eq:massmatrix2} $m$ and $M$ are real diagonal matrices, while
$V = U_\rL^T O_\rR$ is a unitary matrix, $V V^\dag = 1$, so $M'(0,0)$
is Hermitian with real eigenvalues that are the masses of the mass
eigenstates of neutrinos. In general, $M'(0,0)$ may have 15 independent
parameters: $m_i$ and $M_i$ ($i=1$, 2 ,3), while there are three Euler
angles and six phases $V$. Three phases can be absorbed into the
definition of $\nu'_\rL$.

Assuming the hierarchy $m_i \ll M_j$, we can integrate out the 
right-handed (heavy) neutrinos and obtain
an effective higher dimensional operator with Majorana
mass terms for the left-handed neutrinos
\beq
\cL_{\rm dim-5}^\nu =
- \frac12 \sum_{i}
m_{\rM,i} \left(1+\frac{h}{v}\right)^2
\Big(\overline{\nu^{'c}_{i,\rL}} \nu'_{i,\rL} + {\rm h.c.}\Big)
\,.
\eeq
The Majorana masses
\beq
m_{\rM,i} =  \frac{m_i^2}{M_i}
\eeq
are suppressed by the ratios $m_i/M_i$ as compared to $m_i$. The latter
have a similar role in the Lagrangian as the mass parameters of the
charged leptons, so one may assume $m_i \sim$ O(100\,keV), while the masses of
the right-handed neutrinos can be naturally around O(100\,GeV), so that
$m_i/M_i\sim\rO(10^{-6\pm1})$ and $m_{\rM,i} \lesssim 0.1$\,eV.  Thus
if $m_i \ll M_i$, then the {\em mixing between the light and heavy
neutrinos will be very small, the $\nu'_{i,\rL}$ can be considered as
the mass eigenstates that are mixtures of the left-handed weak
eigenstates, and whose masses can be small naturally} as suggested by
phenomenological observations.

As we can only observe neutrinos together with their flavours through
their charged current interactions, it is more natural to use the
flavour eigenstates than the mass eigenstates. In the flavour basis,
the couplings of the leptons to the W boson are diagonal:
\beq
\cL^{(\ell)}_{\rm CC} = -\frac{g_\rL}{\sqrt{2}}
\sum_f \overline{\nu_\rL}^f \slashed{W}^\dag  \ell^f_\rL
+ {\rm h.c.}
\,,
\label{eq:LleptonCC}
\eeq
with summation over the three lepton flavours $f = e$, $\mu$ and
$\tau$. The same charged current interactions in mass basis
$\nu_{\rL,i} = (U_{\rm PMNS})_{if} \nu_\rL^f$,
contains the Pontecorvo-Maki-Nakagawa-Sakata matrix $U_{\rm PMNS}$,
\beq
\cL^{(\ell)}_{\rm CC} = -\frac{g_\rL}{\sqrt{2}} \sum_{i,f=1}^3
\overline{\nu_{\rL,i}}\:(U_{\rm PMNS})_{if}\:\slashed{W}^\dag
\:\ell_{\rL}^f
+ {\rm h.c.}
\,,
\eeq
just like the charged current quark interactions contain the 
Cabibbo-Kobayashi-Maskawa matrix. If the heavy neutrinos are integrated
out, then the matrix $U_\rL$ coincides with the PMNS matrix.  For
propagating degrees of freedom, such as in the case of travelling
neutrinos over macroscopic distances, one should use mass eigenstates
$\nu_{\rL,i}$ and the PMNS matrix becomes the source of neutrino
oscillations in flavour space.  However, in the case of elementary
particle scattering processes involving the left-handed neutrinos, one
can work using the flavour basis, i.e.~with \eqn{eq:LleptonCC} because
the effect of their masses can be neglected. 


\subsection{Re-parametrization into right-handed and mixed couplings}

Having set the $Z$-charges of the matter fields, we can re-parametrize
the couplings to $Z'$ using the new coupling
\beq
g'_{ZY}  = g_Z' - g_Y' = \frac{g_Z - g_Y \sin\tZ}{\cos\tZ}
\,.
\label{eq:gZYdef}
\eeq
Then the covariant derivative in \eqn{eq:cov-dev} becomes
\beq
D^{\mu}_j= \partial^{\mu}
+\ri g_\rL\,\bT\cdot\bom{W}^{\mu}+\ri\,y_{j} g_Y B^{\prime\mu}
+\ri \lp(r_j g_Z' + y_{j} g'_{ZY} \rp)Z^{\prime\mu}
\label{eq:cov-dev2}
\eeq
where $r_j = z_j - y_j$ and its values are given explicitly in the last
column of \tab{tab:charges}. Thus, if a $U(1)_Z$ extension of
$G_{\rm SM}$ is free of gauge and gravity anomalies and the $Z$-charges
of left and right-handed fields are the opposite, then it is equivalent
to a $U(1)_\rR$ extension with tree-level mixed coupling
$g_{ZY}'$ \cite{delAguila:1995rb}, related to the kinetic mixing
parameter $\tZ$ by \eqn{eq:gZYdef}.

Particle phenomenology of the standard model suggests that the
interaction of the fermions through the $Z'$ vector boson must be
suppressed significantly. The origin of such a suppression can be
either a small coupling to $Z'$ or the large mass of $Z'$. Usual
studies in the literature focus on the latter case. Here we explore the
former possibility.

The complete Lagrangian is the sum of the pieces given in
Eqs.~(\ref{eq:LD}), (\ref{eq:LBZW}), (\ref{eq:Lphichi}), (\ref{eq:LY}),
(\ref{eq:nuYukawa}),
\beq
\cL = \cL_\rD + \cL_{B,Z,W} + \cL_{\phi,\chi} + \cL_\rY + \cL_\rY^\nu
\label{eqn:Lfull}
\eeq
with covariant derivative given in \eqn{eq:cov-dev2}, i.e.~the kinetic 
mixing of \eqn{eq:kinetic-mixing} is also taken into account.


\subsection{Mixing in the neutral gauge sector}

The neutral gauge fields of the standard model and the $Z'$
mix, which leads to mass eigenstates $A_\mu$, $Z_\mu$ and $T_\mu$
(not to be confused with the isospin components $T_i$, $i=1$, 2, 3).
The mixing is described by a $3\times 3$ mixing matrix as
\beq
\lp(\!\!\begin{array}{c}
 W^3_\mu \\
 B'_\mu \\
 Z'_\mu
\end{array}\!\!\rp) =
\lp(\!\!\begin{array}{rrr}
 \cos\tW\cos\tT & \cos\tW \sin\tT & \sin\tW \\
-\sin\tW\cos\tT &-\sin\tW \sin\tT & \cos\tW \\
-\sin\tT        & \cos\tT         & 0
\end{array}\!\!\rp) 
\lp(\!\!\begin{array}{c}
 Z_\mu \\
 T_\mu \\
 A_\mu
\end{array}\!\!\rp)
\,.
\eeq
For the Weinberg mixing angle $\tW$ we have the usual value
$\sin\tW = g_Y/\sqrt{g_\rL^2 + g_Y^2}$.  We introduce the notion of
reduced coupling defined by $\gamma_i = g_i/g_\rL$, i.e.~$\gamma_\rL =1$.
Then we have
\beq
\sin\tW = \frac{\gamma_Y}{\sqrt{1+\gamma_Y^2}}
,\qquad
\cos\tW = \frac{1}{\sqrt{1+\gamma_Y^2}}
\label{eq:sW-cW}
\eeq
and for the mixing angle $\tT$ of the $Z'$ boson we find
\beq
\bsp
\sin \tT &= \left[\frac12\left(1
- \frac{1 - \kappa^2 - \tau^2}{\sqrt{(1 + \kappa^2 + \tau^2)^2-4\tau^2}}
\right)\right]^{1/2}
,\\
\cos \tT &= \left[\frac12\left(1
+ \frac{1 - \kappa^2 - \tau^2}{\sqrt{(1 + \kappa^2 + \tau^2)^2-4\tau^2}}
\right)\right]^{1/2}
\,,
\label{eq:sT-cT}
\esp
\eeq
so $\tan(2 \tT) = 2 \kappa/(1 - \kappa^2 - \tau^2)$, with
\beq
\kappa = \frac{\gamma'_{ZY}+\gamma_Z'}{\sqrt{1+\gamma_Y^2}}
\,,\quad
\tau = 2 \frac{\gamma_Z' \tan \beta}{\sqrt{1+\gamma_Y^2}} 
\label{eq:kappa-tau}
\eeq
and
\beq
\tan\beta = \frac{w}{v}
\eeq
is the ratio of the scalar vacuum expectation values (not a scalar
mixing angle).  For small values of the new couplings $\gamma'_{ZY}$
and $\gamma_Z'$, implying small $\kappa$, we have
\beq
\tT = \kappa +\rO(\tau^2,\kappa^3)\,.
\label{eq:thetaT=kappa}
\eeq

The charged current interactions remain the same as in the standard model.
The neutral current Lagrangian can be written in the form
\beq
\cL_{\rm NC} = \cL_{\rm QED} + \cL_{Z} + \cL_{T}
\eeq
where the first term is the usual Lagrangian of QED,
\beq
\cL_{\rm QED} = -e A_\mu J^\mu_{\rm em}
\,,\quad
J^\mu_{\rm em} = 
\sum_{f=1}^{3}\sum_{j=1}^{3} e_j \Big(
  \overline{\psi}^f_{q,j}(x)\gamma^\mu\psi_{q,j}^f(x)
+ \overline{\psi}^f_{l,j}(x)\gamma^\mu\psi_{l,j}^f(x)\Big)
\,,
\label{eq:LQED}
\eeq
the second one is a neutral current coupled to the $\Z0$ boson,
\beq
\cL_{Z} = -e Z_\mu
\Big( \cos\tT J^\mu_{Z} - \sin\tT J^\mu_{T}\Big)
= -e Z_\mu J^\mu_{Z} + \rO(\tT)
\label{eq:LZ0NC}
\eeq
and the third one is the neutral current coupled to the $\T0$ boson,
\beq
\cL_{T} = -e T_\mu
\Big( \sin\tT J^\mu_{Z} + \cos\tT J^\mu_{T}\Big)
= -e T_\mu J^\mu_{T} + \rO(\tT)
\,.
\label{eq:LTNC}
\eeq
In \eqn{eq:LQED} $e$ is the electric charge unit and $e_j$ is the
electric charge of field $\psi_j$ in units of $e$. In
\eqns{eq:LZ0NC}{eq:LTNC} $J^\mu_{Z}$ is the usual neutral
current,
\beq
J^\mu_{Z} = 
\sum_{f=1}^{3}\sum_{j=1}^{3} 
\frac{T_3 - \sin^2\tW\,e_j}{\sin \tW \cos\tW} \Big(
  \overline{\psi}^f_{q,j}(x)\gamma^\mu\psi_{q,j}^f(x)
+ \overline{\psi}^f_{l,j}(x)\gamma^\mu\psi_{l,j}^f(x)\Big)
\,,
\label{eq:JZ0}
\eeq
while the new neutral current has the same dependence on
fermion dynamics with different coupling strength:
\beq
J^\mu_{T} = 
\sum_{f=1}^{3}\sum_{j=1}^{3} 
\frac{\gamma'_Z r_j + \gamma'_{ZY} y_j}{\sin \tW} \Big(
  \overline{\psi}^f_{q,j}(x)\gamma^\mu\psi_{q,j}^f(x)
+ \overline{\psi}^f_{l,j}(x)\gamma^\mu\psi_{l,j}^f(x)\Big)
\,.
\label{eq:JT}
\eeq
We can rewrite these currents as vector--axialvector currents using
the non-chiral fields $\psi_f$ 
\beq
J^\mu_{X} = 
\sum_{f}
  \overline{\psi}_f(x)\gamma^\mu\big(v_f^{(X)} - a_f^{(X)}\gamma_5\big)\psi_f(x)
\,,\quad X = Z\textrm{ or }T
\,,
\label{eq:v-acurrents}
\eeq
with vector couplings $v_f^{(X)}$ and axialvector couplings $a_f^{(X)}$
given in \app{sec:FeynmanRules} and the summation runs over all quark and
lepton flavours. Clearly, the QED current $J^\mu_{\rm em}$ can also be
written using non-chiral fields in the form of \eqn{eq:v-acurrents}
with $v_f^{(\rm em)} = e_f$ and $a_f^{(\rm em)} = 0$.  

As the dependence on the couplings and charges of the neutral currents
in \eqns{eq:JZ0}{eq:JT} are very different for different fermion 
fields, the only way that the standard model phenomenology is not
violated by the extended model if $\tT$ is small, which supports
the expansions used in \eqns{eq:LZ0NC}{eq:LTNC}.

To define the perturbation theory of this model explicitly, we present
the Feynman rules 
in \app{sec:FeynmanRules}.


\subsection{Masses of the gauge bosons}

The photon is massless, while the masses of the massive neutral bosons
are
\beq
M_{Z} = M_{W}\frac{\cos\tT}{\cos\tW}
\Big[ (1+\kappa \tan\tT)^2 + (\tau\tan\tT)^2 \Big]^{1/2}
\label{eq:MZ0}
\eeq
and
\beq
M_{T} = M_{W}\frac{\sin\tT}{\cos\tW}
\Big[ (1-\kappa \cot\tT)^2 + (\tau \cot\tT)^2 \Big]^{1/2}
\label{eq:MT}
\eeq
where $M_W=\frac12 v g_L$ and we assumed $M_{T} < M_{Z}$.
Indeed, in order to have $M_{Z}$ within the experimental uncertainty
of the known measured value, we need $\tT \simeq 0$, which justifies
the expansions at $\kappa=0$,
\beq
M_{Z} = \frac{M_{W}}{\cos\tW}
\lp(1 + \rO(\kappa^2)\rp)
\simeq \frac{M_{W}}{\cos\tW}
\eeq
and
\beq
M_{T} = \frac{M_{W}}{\cos\tW}
\tau\lp(1 + \rO(\kappa^2)\rp)
\simeq M_{Z'}
\eeq
where we used \eqn{eq:thetaT=kappa} and $M_{Z'} = w g_Z'$. Thus $\tau$
can also be written as  the ratio of the masses of the two massive
neutral gauge bosons,
\beq
\tau = \frac{M_{Z'}}{M_W} \cos\tW  \simeq \frac{M_{T}}{M_{Z}}
\,,
\label{eq:tau}
\eeq
justifying our assumption on the hierarchy of masses. In fact, unless
$w \gg v$, we find $M_{T} \ll M_{Z}$.


\subsection{Free parameters}

There are five parameters in the scalar sector, $\lambda_\phi$,
$\lambda_\chi$, $\lambda$, $v$ and $w$ that has to be determined
experimentally, while the values of $\mu_\phi$ and $\mu_\chi$ (at tree
level) are given in \eqn{eq:scalarmasses}.
However, it is more convenient to use parameters that can be measured
more directly, for instance,
\beq
M_h\,,\;
M_H\,,\;
\sin\tS\,,\;
v = (\sqrt{2} G_{\rm F})^{-1/2}\;\mbox{and}\;
\tan\beta\,,
\eeq
of which we know two from measurements: one of the scalar masses and
Fermi's constant.

In addition to the neutrino Yukawa couplings (or neutrino masses and
PMNS mixing parameters), there are five free parameters in the model
that we choose as the mass of the new scalar particle $M_h$ or $M_H$
(the other being fixed by the mass of the Higgs boson), the scalar and
vector mixing angles, the ratio of the vacuum expectation values and
(essentially) the new gauge coupling:
\beq
\sin\tS\,,\;
\sin\tT\,,\;
\tan\beta\,,\;
\tau
\,.
\label{eq:freeparameters}
\eeq
It can be shown \cite{Peli:2019xwv} that requiring stable vacuum up to
the Planck scale, the Higgs particle coincides with the scalar $h$ and
according to a one-loop analysis of the running scalar couplings $M_h$
falls into the range [144,558]\,GeV.

The other parameters can be expressed in terms of the free ones as follows:
$w = v \tan\beta$,
\beq
\bsp
\lambda_\phi &= \frac1{2 v^2} \lp(M_{h/H}^2 \cos^2\tS + M_{H/h}^2 \sin^2 \tS\rp)
\,,\\[2mm]
\lambda_\chi &= \frac1{2 w^2} \lp( M_{H/h}^2 \cos^2\tS + M_{h/H}^2 \sin^2\tS\rp)
\,,\\[2mm]
\lambda &= \sin(2\tS)\,\frac{M_H^2-M_h^2}{2 v w}
\esp
\eeq
(first indices are to be used if $\lambda_\phi v^2 < \lambda_\chi w^2$,
the second ones otherwise) and
\beq
\bsp
\tan\tZ &= \frac{\tau-\kappa \tan\beta}{\tan\beta \sin\tW}
\,,\\[2mm]
\kappa &= \cot(2\tT) \Big(\sqrt{1+(1-\tau^2) \tan^2(2\tT)}-1\Big)
= (1-\tau^2) \sin\tT + \rO(\tT^3)
\,,\\[2mm]
\gamma_Z' &= \frac{\tau}{2 \tan\beta \cos\tW}
\,,\quad
\gamma_Y' = \frac{\tau-\kappa \tan\beta}{\tan\beta \cos\tW}
\,,\quad
\gamma'_{ZY} = \frac{2\kappa \tan\beta-\tau}{2 \tan\beta \cos\tW}
\,.
\label{eq:parameters}
\esp
\eeq


\section{Possible consequences}

Our hope in devising this model is to explain the established 
experimental observations listed in the introduction. We envisage the
following scenario:
\begin{itemize}
\itemsep=-2pt
\item
The massive $\T0$ vector boson is a natural candidate for WIMP dark matter
if it is sufficiently stable, i.e.~its mass is below the threshold of
electron-positron pair production, which requires that the new force
is super-weak, $\tau \sim 10^{-5}$. Such a light vector boson is not
yet excluded by beam-dump experiments (see for instance
\cite{Alexander:2016aln} and references therein). We study the most
recent exclusion limits in \sect{sec:BaBar-N664}. A new technology to
search for electron recoils from the interaction of sub-GeV dark matter
particles with electrons in silicon start to become sensitive to dark
matter searches of mass as low as about 500\,keV \cite{Crisler:2018gci}.
\item
Majorana neutrino mass terms for the right-handed neutrinos and
Yukawa interactions between the left- and right-handed neutrinos and the
BEH vacuum are generated by the spontaneous symmetry breaking of the
scalar fields as outlined in \sect{ssec:Yukawas}.  This scenario 
provides a possible origin of neutrino oscillations and effective
Majorana mass terms for the left-handed neutrinos.
\item
The neutrino Yukawa terms provide a source for the PMNS matrix as shown in
in \sect{ssec:Yukawas}, which in turn can produce leptogenesis (and hence 
baryogenesis).
\item
The vacuum of the $\chi$ scalar has a charge $z_j=-1$ (or $r_j = -1$) that
may be a source of the current accelerated expansion of the universe.
\item
The second scalar together with the established BEH field can cause
hybrid inflation.
\end{itemize}
In order that the model makes these explanations credible, we have to
find answer to the following question: {\em Is there any region of the
parameter space of the model that is not excluded by experimental
results, both established in standard model phenomenology and elsewhere?}
Of course, answering this question requires studies well beyond the
scope of a single article. Here we shall focus on the constraints over
the parameter space that can be obtained from the standard model 
phenomenology and in particular from the anomalous magnetic moment
of the muon and searches for light neutral vector boson.

\section{Anomalous magnetic moment of the muon}
\label{sec:muong-2}

There is a long standing deviation between the experimental result and
predicted standard model value of the anomalous magnetic moment of the
muon \cite{Bennett:2006fi},
\beq
a_{\mu}^{(\rm exp)} - a_{\mu}^{(\rm SM)} = 268(76)\cdot 10^{-11}
\,.
\label{eq:amuexp}
\eeq
Here we assume that this difference--which will be tested by the 
improved precision of future experiments--is due to the effect of
the new gauge boson to the anomalous magnetic moment and we estimate the
allowed values for the ratio $\tan\beta$ of the vacuum expectation
values and that of the mixed coupling $\gamma_{ZY}'$ and the right
coupling
$\gamma_Z'$, 
\beq
\rho'_Z = \frac{\gamma_{ZY}'}{\gamma_Z'}
= 1-\frac{\gamma_{Y}'}{\gamma_Z'} 
= 1-\epsilon \frac{g_{Y}}{g_Z}
\,.
\label{eq:rhoZ'}
\eeq
Note that if $\rho'_Z$ were vanishing, then the new gauge boson couples
only to right-handed fermions, while $\rho'_Z = 1$ implies vanishing
kinetic mixing when $\tau = \kappa/\tan\beta$, so the number of free 
parameters reduces by one.  

As the new $U(1)_Z$ sector may influence the standard model
phenomenology only within the current experimental uncertainties, the
new gauge coupling must be small. Therefore, the use of first order
perturbation theory is justified. At one-loop accuracy, the only new
contributions to the anomaly constant $a_\mu = (g_\mu-2)/2$ emerge from
to the modified $Z\bar{\mu}\mu$ interaction and the new interaction
$T\bar{\mu}\mu$, both presented in the Appendix. The only new Feynman
graph is a triangle with the exchange of a $\T0$ boson between the muon
legs, which is formally identical to the triangle with the exchange of
a $\Z0$ boson between the muon legs as shown in \fig{fig:muon-graph}.
There is also a graph with the exchange of the new scalar, but that is
suppressed by the super-weak new gauge coupling as compared to the
(also negligible) contribution from the exchange of the Higgs boson.
Consequently, the computation follows the same steps as in the case of
the electroweak corrections
\cite{Jackiw:1972jz,Bars:1972pe,Bardeen:1972vi,Fujikawa:1972fe},
so we present only the result for the exchange of a massive U(1) gauge
boson $X$ $(X = \Z0)$  or $\T0$): 
\begin{figure}
\begin{center}
\includegraphics[width=0.2\textwidth]{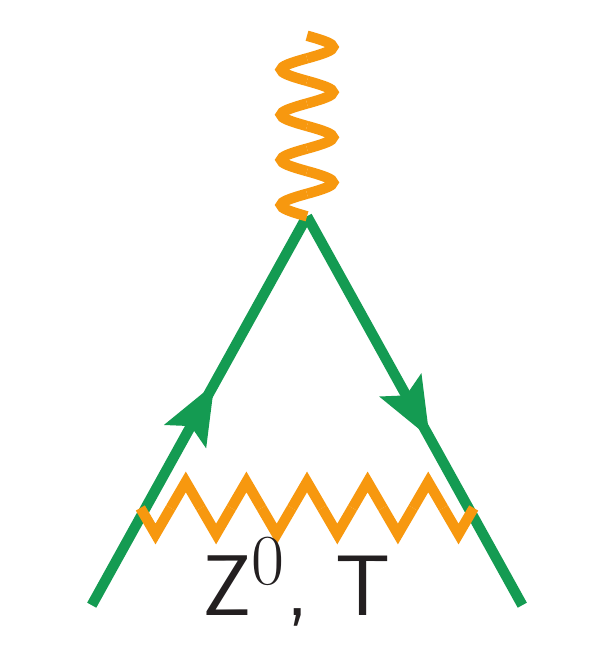}
\end{center}
\caption{\label{fig:muon-graph}
Feynman diagram containing the effect of the new vector boson on
the anomalous magnetic moment of the muon at one loop accuracy.
}
\end{figure}
\beq
a^{(X)}_\mu (h_f,\tT) = \frac{G_{\rm F} m_\mu^2}{6\sqrt{2} \pi^2}
\Big[3 C_X^+ C_X^- - (C_X^+)^2 - (C_X^-)^2\Big]
\,,
\label{eq:amuZ}
\eeq
with coefficients $C_X^+$ and $C_X^-$ given for both gauge bosons in
the Appendix in terms of flavour dependent constants $g_f^\pm$ and
$h_f^\pm$ defined in \eqn{eq:gfs}.  For the muon
\beq
g_\mu^+ = \frac{\sW}{\cW}
\,,\quad
g_\mu^- = \frac{\sW^2 - \frac12}{\sW\cW}
\,,\quad
h_\mu^- = -\frac{\gamma_Z'}{2\sW}
\,,\quad
h_\mu^+ = -\frac{\gamma_Z'}{2\sW}\,(1+2\rho_Z')
\,.
\eeq
The contribution of the $\Z0$ boson in the standard model is recovered
by setting $h_f^\pm = 0$ and $\tT = 0$. Thus, the complete new
contribution to the $a_\mu$ in this model is given by
\beq
\Delta a_\mu = a^{\rm (T+SM)}_\mu - a^{\rm (SM)}_\mu =
a^{(\Z0)}_\mu(h_f,\tT) - a^{(\Z0)}_\mu(0,0) + a^{(\T0)}_\mu(h_f,\tT)
\,.
\eeq

As mentioned before, the standard model phenomenology requires
$\tT\simeq 0$, which justifies the expansion in $\tT$:
\beq
\bsp
\Delta a_\mu = \frac{G_{\rm F} m_\mu^2}{6\sqrt{2} \pi^2}
\bigg(\frac{(1 + \rho_Z')\cos^2\tW-\frac12}{\tan\beta}
+ \rO(\tT,\:\gamma_Z')\bigg)^2
\label{eq:Damu}
\esp
\eeq
where we used \eqns{eq:MZ0}{eq:MT} together with \eqn{eq:thetaT=kappa}
and the definitions in \eqns{eq:sT-cT}{eq:kappa-tau}. According to
\Ref{Tanabashi:2018oca}, numerically
\beq
\frac{G_{\rm F} m_\mu^2}{6\sqrt{2} \pi^2} \simeq 155.5\cdot 10^{-11}
\,.
\label{eq:factor}
\eeq

\begin{figure}
\begin{center}
\includegraphics[width=0.5\textwidth]{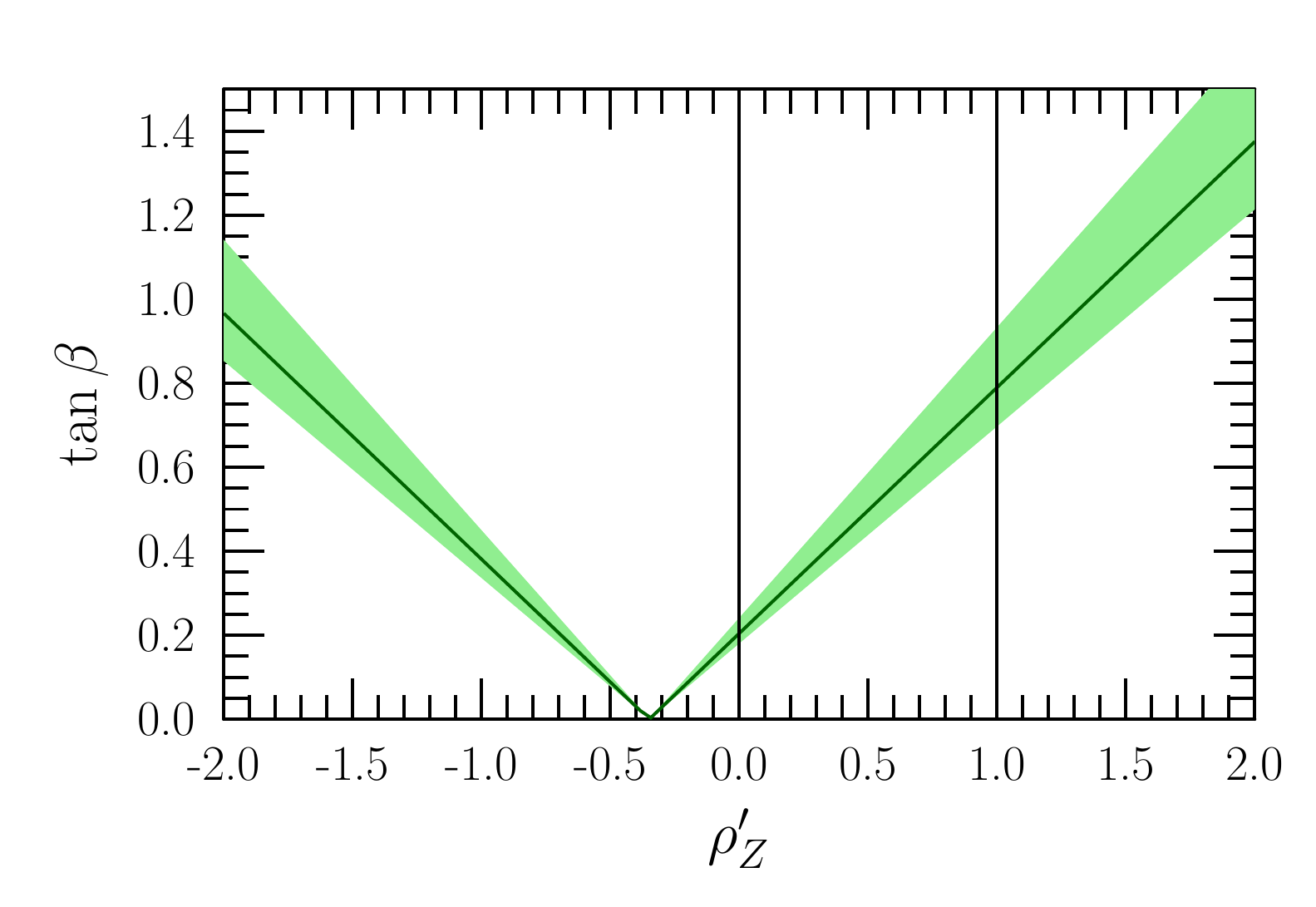}
\end{center}
\caption{\label{fig:rho-tanbeta}
Allowed region of parameter space assuming that the $\T0$ boson is
responsible for the deviation between the measured value of the 
anomalous magnetic moment of the muon and the prediction for it
in the standard model. The band represents the uncertainty derived
from the experimental uncertainty of the measurements.
}
\end{figure}

The deviation in \eqn{eq:amuexp} is explained by the contribution in
\eqn{eq:Damu} if $\rho_Z'$ and $\tan\beta$ are confined to the region
determined by 
\beq
\tan\beta \simeq
\sqrt{\frac{155.5}{268\pm 76}} \Big( 0.76848(1 + \rho_Z') - 0.5 \Big)
\,,
\eeq
and shown in \fig{fig:rho-tanbeta}. We see that smaller difference between
the data and the standard model prediction essentially implies larger
$\tan\beta$, i.e.~larger $w$.

\section{Exclusion limits from searches for invisibly decaying
light neutral vector boson}
\label{sec:BaBar-N664}

There is a long list of experiments devised to search directly for
light neutral vector boson. The results of the most recent ones put
severe constraint in the plane of kinetic mixing parameter 
$\epsilon$ and mass of the boson $M_T$
\cite{Lees:2017lec,NA64:2019imj} in the framework when the new gauge
boson couples only by the kinetic mixing term to the fields of the
standard model, which means that its coupling to the fermions is purely
vector like.  It is interesting to test whether the parameter space
allowed in \fig{fig:rho-tanbeta}, i.e.~the region favoured by the muon
magnetic moment anomaly at present, has any overlap with the still
allowed region in these direct search experiments.  As in the model
presented here the $\T0$ boson has both $v-a$ couplings to the fermions
(see \eqn{eq:v-acurrents}), one has to derive an effective kinetic
mixing parameter $\epeff$ that can be used in vector like interactions.

\begin{figure}
\begin{center}
\includegraphics[width=0.4\textwidth]{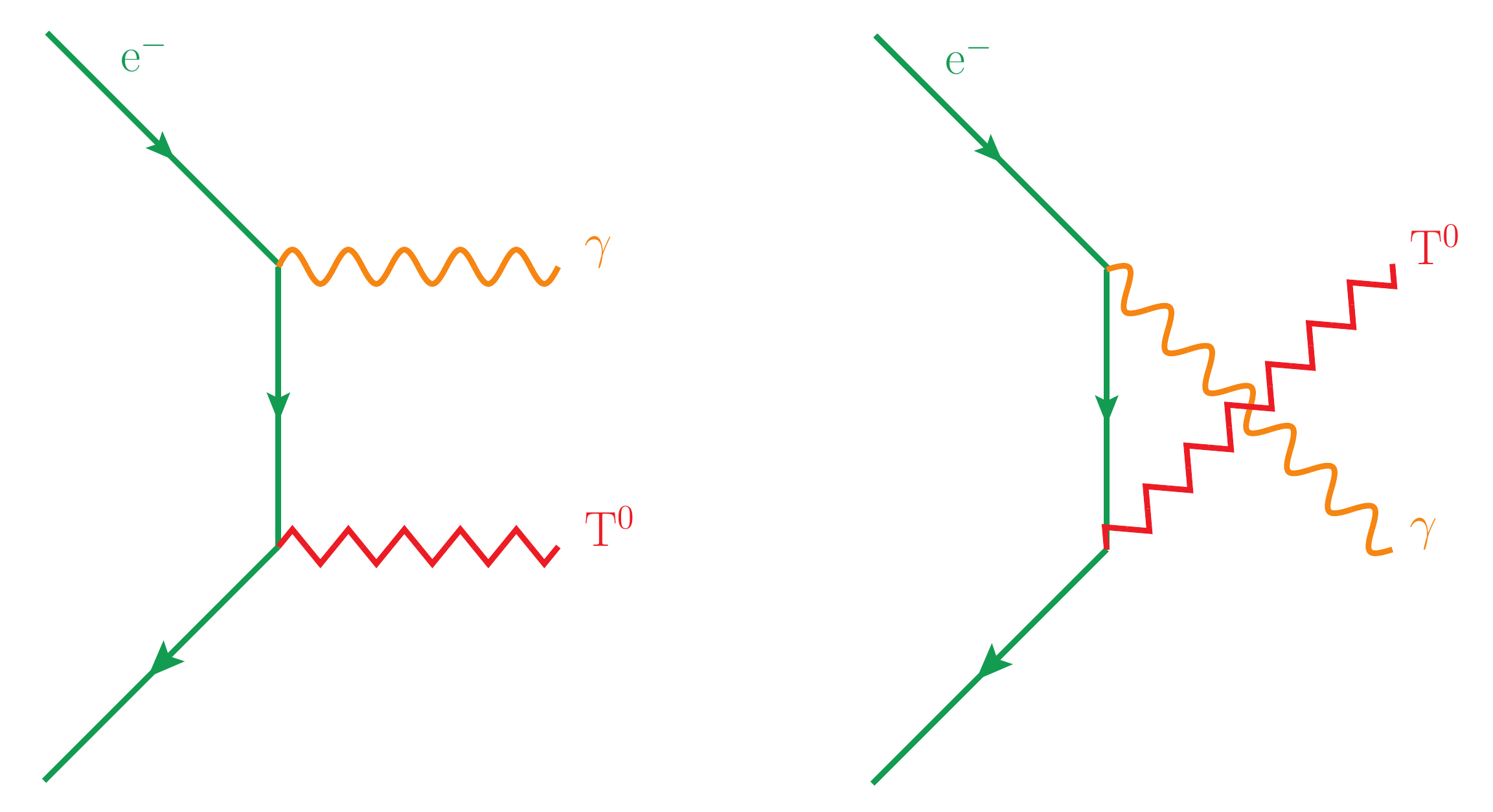}
\end{center}
\caption{\label{fig:BaBar}
Feynman diagrams producing a photon and a $\T0$ boson in electron-positron
annihilation.
}
\end{figure}

In the case of the BaBar experiment the search channel is an associated
production of a light neutral vector boson with a photon in electron-positron
annihilation \cite{Lees:2017lec}. The massive boson is assumed
to decay invisibly to the detector, hence the signal is a single photon
plus missing energy and momentum. At lowest order in perturbation theory
the production channel is given by the diagrams shown in \fig{fig:BaBar}.
The corresponding cross section,
\beq
\sigma(\re^+\re^- \to \gamma A') =
\frac1{2s} \int\!\rd\phi_2 |\cM_{\re^+\re^- \to \gamma A'}|^2
\,,
\eeq
is proportional to the square of the kinetic mixing parameter $\epsilon$.
As the $\T0$ boson couples to the fermions with $v-a$ type couplings, the 
cross section for the $\re^+\re^- \to \gamma \T0$ process,
\beq
\sigma(\re^+\re^- \to \gamma \T0) =
\frac1{2s} \int\!\rd\phi_2 |\cM_{\re^+\re^- \to \gamma \T0}|^2
\,,
\eeq
is proportional to $\Big(v_\re^{(T)}\Big)^2 + \Big(a_\re^{(T)}\Big)^2$. 
Hence, we can define the effective kinetic mixing parameter for the 
$\re^+\re^- \to \gamma \T0$ production channel as
\beq
\epeff = \sqrt{
\frac{\sigma(\re^+\re^- \to \gamma \T0)}
{\sigma(\re^+\re^- \to \gamma A')/\epsilon^2}
} = \sqrt{\Big(v_\re^{(T)}\Big)^2 + \Big(a_\re^{(T)}\Big)^2}
= \frac{\gamma_Z'}{2\sW}\sqrt{\frac52 \rho_Z'{}^2+\rho_Z'+\frac12}
\,.
\eeq

In the case of the NA64 experiment the production channel corresponds
to the crossing of the BaBar production channel \cite{NA64:2019imj},
$(\re^+\re^- \to \gamma A') \longrightarrow (\re^- \gamma^* \to \re^- A')$
where the virtual photon emerges from the nucleus on which the electron
scatters.  As a result, the effective kinetic mixing parameter is the
same as in the BaBar experiment.

\begin{figure}
\begin{center}
\includegraphics[width=0.7\textwidth]{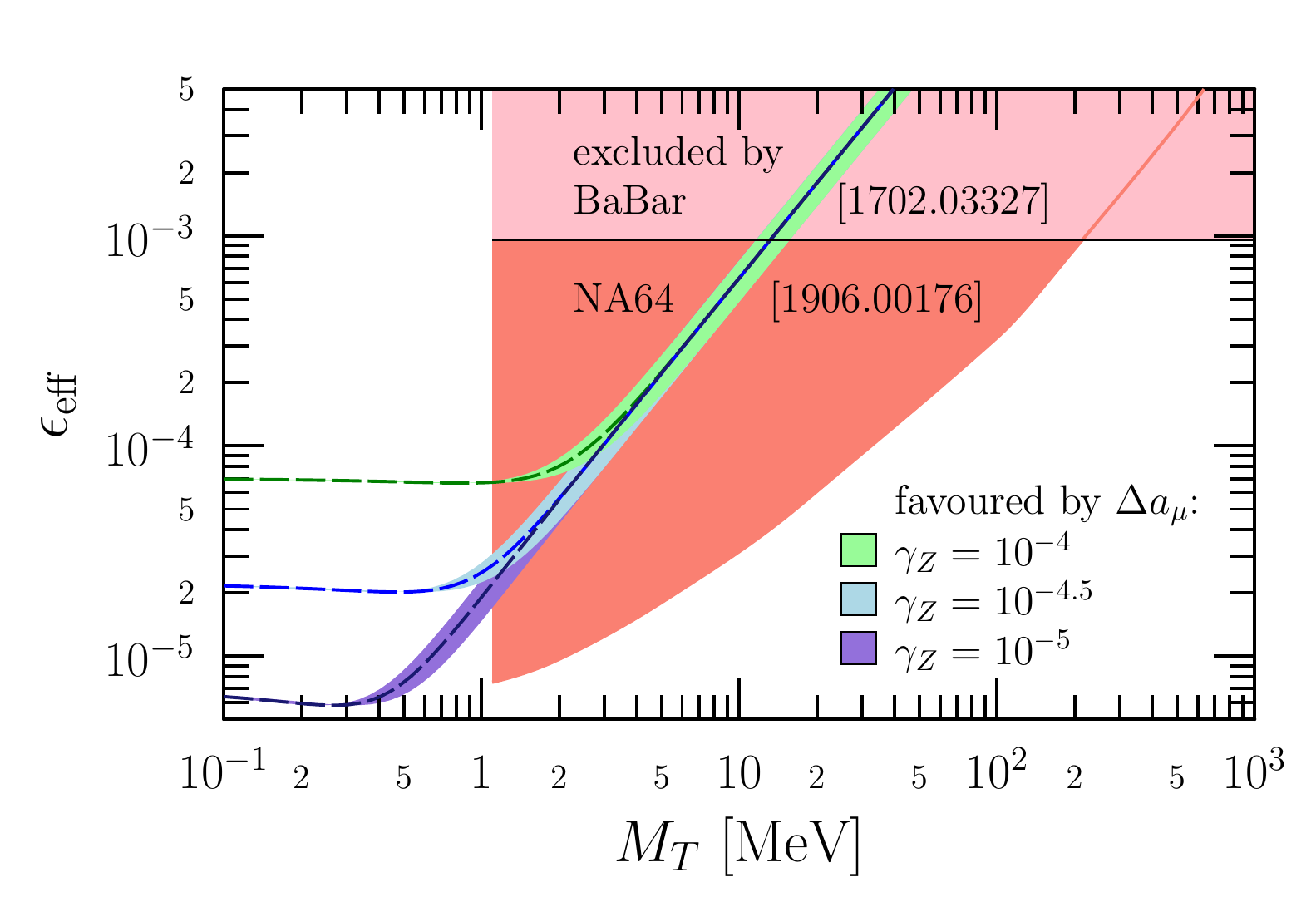}
\end{center}
\vspace*{-10pt}
\caption{\label{fig:eps-mT}
Region in the $\epeff - M_T$ plane favoured by the muon magnetic moment
anomaly at several values of the $\gamma_Z'$ coupling vs.~the exclusion
limits set by the BaBar \cite{NA64:2019imj} and NA64 \cite{Lees:2017lec}
experiments.}
\end{figure}

Using Eqs.~(\ref{eq:sW-cW}), (\ref{eq:kappa-tau}) and (\ref{eq:tau}),
we can translate the preferred region in \fig{fig:rho-tanbeta} into the
preferred region in the $\epeff - M_T$ plane, shown in \fig{fig:eps-mT}.
We see that the direct searches for the light neutral gauge boson allows
for the interpretation of the deviation 
$a_{\mu}^{(\rm exp)} - a_{\mu}^{(\rm SM)} = 268(76)\cdot 10^{-11}$
with the existence of the $\T0$ boson only if it has mass below the 
electron-pair threshold, which supports our
previous assumption about the smallness of the mass $M_T$, and
consequently of the coupling $g_Z'$, i.e.~the new force is super weak.

\section{Conclusions}

In this paper we collected the well established experimental
observations that cannot be explained by the standard model of particle
interactions. We have then proposed an anomaly free extension by a
$U(1)_Z$ gauge group, which is the simplest possible model. We also
assumed the existence of a new complex scalar field with $Z$-charge
only (i.e.~neutral with respect to the standard model interactions) and
three right-handed neutrinos.
In order to fix the $Z$-charges of the particle spectrum we 
assumed that the left- and right-handed neutrinos have opposite
$Z$-charges. Thus such a model predicts the existence of (i) a massive
neutral vector boson, (ii) a massive scalar particle and (iii) three
massive right-handed neutrinos. The left-handed neutrinos remain
massless as in the standard model, but their Yukawa interactions
with the BEH field and the right-handed neutrinos provide a field
theoretical basis for explaining neutrino oscillations and predict
effective Majorana masses for the propagating mass eigenstates.

We have discussed how the new neutral gauge field $Z^\mu$ mixes with
those of the standard model ($B^\mu$ and $W_3^\mu$) and argued that the
mixing results in a new vector boson $\T0$ of a small mass related to
the small new gauge coupling and small mixing with the standard model
vector fields. We also presented the Feynman rules of the model in
unitary gauge and collected the new free parameters.

In order that the predictions of the model be credible, we have to
answer whether there is any region of the parameter space
that is not excluded by experimental results established in standard
model phenomenology or elsewhere.  To answer such a question with
satisfaction, studies well beyond the scope of a single article are
needed, which forecasts an exciting research project. As a first
check, we computed the contribution of the new vector boson to the
anomalous magnetic moment of the muon and used the difference
$\Delta a_\mu$ between the prediction of the standard model and the
measured values to constrain the parameter space of the model.
Comparing the allowed parameter space to the exclusion limits set by
direct searches for invisible decays of dark photons by the NA64
experiment, we found that the present value of $\Delta a_\mu$ allows
for the existence of the $\T0$ boson if its mass is below the threshold
of electron-positron pair production. An analysis of the ultraviolet
behaviour of the scalar couplings is presented in \Ref{Peli:2019xwv}.  

\subtitle{Acknowledgments}

I am grateful G.~Cynolter, D.~Horv\'ath, S.~Iwamoto, A.~Kardos and S.~Katz
for their constructive criticism on the manuscript. This work was
supported by grant K 125105 of the National Research, Development and
Innovation Fund in Hungary.  


\providecommand{\href}[2]{#2}\begingroup\raggedright\endgroup
\appendix

\section{Feynman rules}
\label{sec:FeynmanRules}

The Feynman rules of the model are obtained from the complete Lagrangian
in \eqn{eqn:Lfull}. For studying the UV behaviour of the model, it is 
convenient to use the Feynman rules before SSB, while for low energy
phenomenology the rules after SSB are needed. In this paper we concentrated
only on a simple application of the latter that did not require
renormalization, so rules in the unitary gauge were sufficient. The
propagators of the new fields are related trivially to those of the
standard fields. Thus, we present only the vertices, neglecting the
rules related to QCD, which are unchanged.


\subsection*{Feynman rules after SSB}

We present the rules in unitary gauge.
\begin{itemize}
\item Gauge field interactions:
\begin{itemize}
\item The cubic gauge field interactions of fields $V_{1,\alpha}$,
$V_{2,\beta}$ and $V_{3,\gamma}$ with all-incoming kinematics,
$p^\mu+q^\mu+r^\mu=0$ are
$\Gamma_{\alpha,\,\beta,\,\gamma}\lp(p,q,r\rp)= 
\ri e C V_{\alpha,\,\beta,\,\gamma}\lp(p,q,r\rp)$
where 
\[V_{\alpha,\,\beta,\,\gamma}\lp(p,q,r\rp)= 
\lp(p-q\rp)_{\gamma}g_{\alpha\beta} 
+\lp(q-r\rp)_{\alpha}g_{\beta\gamma} 
+\lp(r-p\rp)_{\beta}g_{\alpha\gamma}\,,
\]
while $C$ depends on the type of the gauge bosons participating in the
interaction as follows
\[
\begin{array}{l|l}
\hline \hline
V_1 V_2 V_3     & C \\
\hline
\gamma W^+ W^- & 1 \\
Z W^+ W^- & \displaystyle{\frac{\cW}{\sW}}\cT \\[4mm]
T W^+ W^- & \displaystyle{\frac{\cW}{\sW}}\sT \\
\hline \hline
\end{array}
\]
\item The quartic gauge field interactions of fields
$V_{1,\alpha}$, $V_{2,\beta}$, $V_{3,\gamma}$ and $V_{4,\delta}$ are\\
$\Gamma_{\alpha,\,\beta,\,\gamma,\,\delta} = \ri e^2 C\,
\lp[2 g_{\alpha\beta}g_{\gamma\delta}
- g_{\alpha\gamma}g_{\beta\delta}
- g_{\alpha\delta}g_{\beta\gamma}\rp]$ where $C$ again depends
on the type of the gauge bosons participating in the interaction
as follows
\[
\begin{array}{l|l}
\hline \hline
V_1 V_2 V_3 V_4 & C \\
\hline
W^+ W^- \gamma \gamma & -1 \\
W^+ W^- \gamma Z & -\displaystyle{\frac{\cW}{\sW}}\cT \\[4mm]
W^+ W^- \gamma T & -\displaystyle{\frac{\cW}{\sW}}\sT \\[4mm]
W^+ W^- Z Z & -\lp(\displaystyle{\frac{\cW}{\sW}}\cT\rp)^2 \\[4mm]
W^+ W^- T Z & -\lp(\displaystyle{\frac{\cW}{\sW}}\rp)^2\cT\sT \\[4mm]
W^+ W^- T T & -\lp(\displaystyle{\frac{\cW}{\sW}}\sT\rp)^2 \\[4mm]
W^+ W^+ W^- W^- & \displaystyle{\frac{1}{(\sW)^2}} \\
\hline \hline
\end{array}
\]
\end{itemize}
\item Scalar interactions:
We denote the standard model Higgs boson by $\cH$, while
the new one by $\cS$. 
\begin{itemize}
\item Cubic scalar interactions can be either of the form
$\ri e \frac{C}{3!} S^3$ where $C$ depends on the type of the scalar
boson participating in the interaction:
\[
\begin{array}{l|l}
\hline \hline
S S S & \qquad C \\
\hline
\cH \cH \cH & \displaystyle{-\frac32\frac{M_h^2 \cos^2\tS + M_H^2\sin^2 \tS}{\sW M_W}}\\[4mm]
\cS \cS \cS & \displaystyle{-\frac32\frac{M_h^2 \sin^2\tS + M_H^2\cos^2 \tS}{\sW M_W \tan\beta}}\\
\hline \hline
\end{array}
\]
or of the form $\ri e \frac{C}{2!} S S S'$
where $C$ depends on the type of the $S$ boson participating in the
interaction: 
\[
\begin{array}{l|l}
\hline \hline
S S S' & \qquad C \\
\hline
\cH \cH \cS & \displaystyle{-\sin\tS \cos\tS \frac{M_H^2 - M_h^2}{2 \sW M_W}}\\[4mm]
\cS \cS \cH & \displaystyle{-\sin\tS \cos\tS \frac{M_H^2 - M_h^2}{2 \sW M_W \tan\beta}}\\
\hline \hline
\end{array}
\]
Recall that $M_{H/h}$ is the mass of the heavier/lighter scalar.
\item The quartic scalar interactions are either of the form
$\ri e^2 \frac{C}{4!}S^4$ where $C$ depends on the type of the scalar
bosons participating in the interaction as follows 
\[
\begin{array}{l|l}
\hline \hline
S S S S & \qquad C \\
\hline
\cH \cH \cH \cH & \displaystyle{-\frac34\frac{M_h^2 \cos^2\tS + M_H^2\sin^2 \tS}{(\sW M_W)^2}}\\[4mm]
\cS \cS \cS \cS & \displaystyle{-\frac34\frac{M_h^2 \sin^2\tS + M_h^2\cos^2 \tS}{(\sW M_W \tan\beta)^2}}\\
\hline \hline
\end{array}
\]
or of the form $\ri e^2 \frac{C}{2!\,2!}\cH^2 \cS^2$ where
\[
C = -\frac34\frac{M_h^2 - M_h^2}{(\sW M_W)^2 \tan\beta}\,.
\]
\end{itemize}
\item Mixed gauge field-scalar interactions:
\begin{itemize}
\item The cubic gauge field-scalar interactions of fields $V_{1,\alpha}$,
$V_{2,\beta}$ and $S$ are
$\ri e g_{\alpha\beta} C$ where $C$ depends on the types of the fields
participating in the interaction as follows\\
\[
\begin{array}{ll}
\hline \hline
V_1 V_2 S & C \\
\hline
W^+ W^- \cH & \displaystyle{\frac{M_W}{\sW}} \\[4mm]
Z Z \cH & \displaystyle{\frac{M_W}{\sW}\frac{(\cT+\kappa\sT)^2}{(\cW)^2}} \\[4mm]
T T \cH & \displaystyle{\frac{M_W}{\sW}\frac{(\sT-\kappa\cT)^2}{(\cW)^2}} \\[4mm]
T Z \cH & \displaystyle{\frac{M_W}{\sW}\frac{(\sT-\kappa\cT)(\cT+\kappa\sT)}{(\cW)^2}} \\[4mm]
Z Z \cS & \displaystyle{\frac{M_W}{\sW\tan\beta}\frac{(\tau \sT)^2}{(\cW)^2}} \\[4mm]
T T \cS & \displaystyle{\frac{M_W}{\sW\tan\beta}\frac{(\tau \cT)^2}{(\cW)^2}} \\[4mm]
T Z \cS & \displaystyle{-\frac{M_W}{\sW}}\frac{\tau^2 \sT \cT}{(\cW)^2} \\[4mm]
\hline \hline
\end{array}
\]
\item Quartic gauge field-scalar interactions $V_\alpha V_\beta SS:$
$\ri e^2 g_{\alpha\beta} C$ where $C$ depends on the type of the gauge
boson participating in the interaction as follows
\[
\begin{array}{l|l}
\hline \hline
V_1V_2SS & C \\
\hline
W^+ W^- \cH \cH & \displaystyle{\frac{1}{2(\sW)^2}} \\[4mm]
Z Z \cH \cH & \displaystyle{\frac{(\cT+\kappa \sT)^2}{2(\cW \sW)^2}} \\[4mm]
T T \cH \cH & \displaystyle{\frac{(\sT-\kappa \cT)^2}{2(\cW \sW)^2}} \\[4mm]
T Z \cH \cH & \displaystyle{\frac{(\sT-\kappa \cT)(\cT+\kappa \sT)}{2(\cW \sW)^2}} \\[4mm]
Z Z \cS \cS & \displaystyle{\frac{(\tau \sT)^2}{2(\cW \sW \tan\beta)^2}} \\[4mm]
T T \cS \cS & \displaystyle{\frac{(\tau \cT)^2}{2(\cW \sW \tan\beta)^2}} \\[4mm]
T Z \cS \cS & \displaystyle{-\frac{\tau^2 \sT \cT}{2(\cW \sW \tan\beta)^2}} \\
\hline \hline
\end{array}
\]
\end{itemize}
\item Gauge field-fermion interactions $V_\alpha \bar{f}_i f_j$:
$-\ri e \gamma_\alpha (C^-P_- + C^+P_+)$ where $C^\pm$ depend on the
type of the gauge boson participating in the interaction, the flavour
$f$ of fermions and family number $i$ and $j$ as follows
\[
\begin{array}{l|l|l}
\hline \hline
V\bar{f}_if_j & C^+ & C^- \\
\hline
\gamma \bar{f}_if_j  & e_f \delta_{ij}  & e_f \delta_{ij} \\
     Z \bar{f}_if_j  & (g_f^+ \cT - h_f^+ \sT) \delta_{ij}& (g_f^- \cT - h_f^- \sT) \delta_{ij}\\
     T \bar{f}_if_j  & (g_f^+ \sT + h_f^+ \cT) \delta_{ij}& (g_f^- \sT + h_f^- \cT) \delta_{ij}\\
   W^+ \bar{u}_id_j  & 0 & \displaystyle{\frac{1}{\sqrt{2} \sW}} V_{ij}\\[4mm]
   W^- \bar{d}_ju_i  & 0 & \displaystyle{\frac{1}{\sqrt{2} \sW}} V^\dag_{ij}\\[4mm]
   W^+ \bar{\nu}_i\ell_j  & 0 & \displaystyle{\frac{1}{\sqrt{2} \sW}} \delta_{ij}\\[4mm]
   W^- \bar{\ell}_j\nu_i  & 0 & \displaystyle{\frac{1}{\sqrt{2} \sW}} \delta_{ij}\\
\hline \hline
\end{array}
\]
where
\beq
g_f^+ = -\frac{\sW}{\cW} e_f
\,,\quad
g_f^- = \frac{T^3_f-\sW^2 e_f}{\sW\cW}
\,,\quad
h_f^\pm = \frac{\gamma_Z' R_f^\pm + \gamma_{ZY}' (e_f-R_f^\mp)}{\sW}
\label{eq:gfs}
\eeq
where $R_f^+ = 1/2$ for $U^f$ or $\nu^f$, $R_f^+ = -1/2$ for $D^f$ or
$\ell^f$ and $R_f^- = 0$.  The vector and axial vector couplings of the
$\Z0$ boson read as
\[
\bsp
&
v^{(Z)}_f =
\frac12\Big(g_f^-+g_f^+\Big) \cT
- \frac12\Big(h_f^-+h_f^+\Big) \sT
\\[2mm] &\quad
= \frac{\Big(T_f^3-2(\sW)^2 e_f\Big) \cT
- \gamma_Z' \Big(2 \rho_Z' e_f + (1 - \rho_Z') r_f\Big)\cW \sT}{2\sW\cW}\,,
\\[2mm] &\quad
= \frac{T_f^3-2(\sW)^2 e_f}{2\sW\cW} + \rO(\tT)
\,, \\[2mm]  &
a^{(Z)}_f =
\frac12\Big(g_f^--g_f^+\Big) \cT
- \frac12\Big(h_f^--h_f^+\Big) \sT
\\[2mm] &\quad
= \frac{T_f^3 \cT + \gamma_Z' (1 - \rho_Z') r_f \cW \sT}{2\sW\cW}
= \frac{T_f^3}{2\sW\cW} + \rO(\tT)
\,,
\esp
\]
while those of the $\T0$ boson are
\beq
\bsp
&
v^{(T)}_f =
\\[2mm] &\quad
= \frac{\Big(T_f^3-2(\sW)^2 e_f\Big) \sT
+ \gamma_Z' \Big(2 \rho_Z' e_f + (1 - \rho_Z') r_f\Big)\cW \cT}{2\sW\cW}
\\[2mm]  &\quad
= \frac{\gamma_Z' \Big(2 \rho_Z' e_f + (1 - \rho_Z') r_f\Big)\cW}{2\sW\cW}
+ \rO(\tT)
\,, \\[2mm]  &
a^{(T)}_f =
\frac{T_f^3 \sT - \gamma_Z' (1 - \rho_Z') r_f\cW  \cT}{2\sW\cW}
\\[2mm] &\quad
= -\frac{\gamma_Z' (1 - \rho_Z') r_f \cW}{2\sW\cW}
+ \rO(\tT)
\,,
\label{eq:v-acouplings}
\esp
\eeq
with $\rho_Z' = \gamma_{ZY}'/\gamma_Z'$ defined in \eqn{eq:rhoZ'}.
\item $H\bar{f}_i f_j$ vertex: $\ri e C$
where
\[
C = - \delta_{ij} \frac{1}{2\sW} \frac{m_{f,i}}{M_W}\,.
\]
\item $S\overline{\nu^c}_{\rR,i} \nu_{\rR,j}$ vertex: $\ri e C$
where
\[
C = - \delta_{ij} \frac{1}{2\sW \tan\beta} \frac{m_{\nu_\rR,i}}{M_W}\,.
\]
\end{itemize}

\end{document}